\documentclass[12pt]{article}
\usepackage[utf8]{inputenc}
\usepackage[english]{babel}
\usepackage[centertags]{amsmath}
\usepackage{amssymb}
\usepackage{bm}
\usepackage{amsbsy}
\usepackage{graphicx}
\usepackage{appendix}
\usepackage{mathrsfs}
\usepackage{epsfig}
\usepackage{color}
\usepackage{euscript}
\usepackage{ulem}
\usepackage{multirow}
\usepackage{cite}

\definecolor{dgreen}{rgb}{0,0.6,0}

\definecolor{darkblue}{rgb}{0., 0, 1}

\definecolor{purple}{rgb}{0.65,0.,0.78}

\definecolor{orange}{rgb}{0.89,0.42,0.05}

\usepackage{jheppubm}

\newcommand{\nn}{\nonumber}
\newcommand{\be}{\begin{equation}}
\newcommand{\ee}{\end{equation}}
\newcommand{\bea}{\begin{eqnarray}}
\newcommand{\eea}{\end{eqnarray}}

\newcommand{\ff}{\mathfrak{f}}

\newcommand{\fz}{\mathfrak{z}}

\newcommand{\fG}{\mathfrak{G}}

\newcommand{\cV}{{\cal V}}

\numberwithin{equation}{section}

\title{Beta-Function Dependence on Running Coupling in Holographic QCD Models 
}

\author{Irina Ya. Aref'eva$^a$, Ali Hajilou$^a$, Pavel Slepov$^a$ and  Marina Usova$^a$}

\affiliation{$^a$Steklov Mathematical Institute, Russian Academy of  Sciences, \\ Gubkina str. 8, 119991, Moscow, Russia}

\emailAdd{arefeva@mi-ras.ru}
\emailAdd{hajilou@mi-ras.ru}
\emailAdd{slepov@mi-ras.ru}
\emailAdd{usovamk@mi-ras.ru}

 \abstract{We study the dependence of beta-function on running coupling constant in  holographic models supported by Einstein-dilaton-Maxwell action for light and heavy quarks. Although, in the previous paper \cite{AHSU}, we considered different types of the dilaton boundary conditions, but since the behavior of $\beta$-function as a function of running coupling does not depend significantly on the boundary condition, we chose one. The corresponding $\beta$-functions are negative and monotonically decreasing functions, and have jumps on the 1-st order phase transitions for both light and heavy quarks. In addition, we compare our holographic results for $\beta$-function as a function of running coupling with perturbation results that obtained within 2-loop calculations.}


\begin{document}

\maketitle

\newpage

\section{Introduction}
% The strongly coupled regime of gauge theories can be explored using holographic duality that describes the correspondence between a class of strongly coupled field theories and weakly-coupled gravitational theories \cite{Maldacena:1997re,Casalderrey-Solana:2011dxg,Arefeva:2014kyw}. The Renormalization group (RG) flow in QFT can be described using a gravitational model coupled to dilaton field, i.e. holographic RG flow 

The strongly coupled regime of gauge theories can be explored using holographic duality, which describes the correspondence between a class of strongly coupled quantum field theories and weakly-coupled gravitational theories \cite{Maldacena:1997re,Casalderrey-Solana:2011dxg,Arefeva:2014kyw}. The Renormalization Group (RG) flow in QFT can be described using  gravity coupled to a dilaton field, i.e., holographic RG flow\cite{Boonstra:1998mp,Heemskerk:2010hk,Kiritsis:2014kua,Gursoy:2018umf}. 
The geometric description of RG flow in this approach is dual to the gravitational solution with specific asymptotics such that the holographic coordinate $z$ corresponds to the energy scale of the QFT. Thereby, the $\beta$-function has a holographic dual that describes the dependence of the running coupling on the energy scale in QFT \cite{BogSchirkov,Wilson:1973jj,
Callan:1970yg,Symanzik:1970rt}.\\

The structure of the QCD phase diagram in the (chemical potential, temperature)-plane is one of the important tasks for high-energy experiments at LHC, RHIC, NICA, and FAIR. Since standard calculations such as perturbation theory do not work in the strongly coupled regime of QCD, we need a non-perturbative approach to describe the physics of strongly interacting quark-gluon plasma (QGP), produced in heavy ion collisions (HIC) at RHIC, LHC, and NICA \cite{Casalderrey-Solana:2011dxg,Arefeva:2014kyw}. 
\\

It is expected from lattice calculations \cite{Brown:1990ev,Philipsen:2016hkv} and some effective phenomenological models \cite{Fu:2019hdw}, that the structure of the QCD phase diagram is different for heavy and light quarks due to its significant dependence on quark masses. 
In light of the significant dependence of lattice calculations on the mass of quarks, holographic QCD serves as a non-perturbative theory required to reproduce the known features of lattice calculations, employing different models for heavy and light quarks.\cite{Andreev:2006ct,Arefeva:2016rob,Yang:2015aia,Arefeva:2018hyo,Li:2017tdz,Arefeva:2020byn,Arefeva:2022bhx, Arefeva:2021mag}. 
Holographic QCD, generally speaking, gives different behavior of $\beta$-functions in different phases, namely, the quark confinement and QGP phases, as well as near the critical lines for heavy and light quarks.
\\

In the context of QCD applications, the holographic RG flow have been widely studied in \cite{Gursoy:2007cb, Gursoy:2007er,Kiritsis:2014kua,Gursoy:2018umf}.
The  RG flows for anisotropic QGP with non-zero chemical potential and temperature are  investigated in \cite{Arefeva:2018hyo,Arefeva:2020aan,Arefeva:2019qen}.
The exact holographic RG flows introduced in  higher-dimensional \cite{Arefeva:2018jyu} and low-dimensional cases
\cite{Golubtsova:2022hfk,Arkhipova:2024iem}.  
\\

The beta-function encodes the dependence of running coupling $\alpha$ on the energy scale of the physical system and is given by
\be
\beta_{QFT}(\alpha)=\frac {\partial \alpha(\Lambda)}{\partial \ln(\Lambda )}~,
\ee 
where $\alpha=\alpha(\Lambda)$ is the running coupling and  $\Lambda$ denotes the energy scale in QFT and related with the holographic coordinate $z$ via $z\sim \frac{1}{\Lambda}$ \cite{Peet:1998wn,Casalderrey-Solana:2011dxg}. In this paper, our goal is to study the $\beta$-function as a function of running coupling constant via holography.
The holographic $\beta$-function is defined  by \cite{deBoer:1999tgo, Gursoy:2007cb,He:2010ye,Arefeva:2019qen,Arefeva:2018jyu,Arefeva:2020aan}
 \be \label{beta-z}
\beta(\alpha)= \,\frac{\partial \alpha} {\partial \, ln B}\,,
 \ee
where $\alpha(z)=e^{\varphi(z)}$, $\varphi$ is the dilaton field, and $B(z)=e^{A(z)}/z$ plays the role of  the energy scale in QFT and  specifies the warp-factor in the holographic metric 
(see details in the text, in particular, \eqref{metric} ). 
 \\

The main goal of this paper is to study the $\beta$-function dependence on the running coupling constant for light and heavy quark holographic models. In the previous paper \cite{AHSU}, we studied the behavior of the $\beta$-function as a function of the holographic coordinate $z$ for different temperatures and chemical potentials. In this research, we focus on the $\beta(\alpha)$ dependence and compare our results with perturbative calculations.
\\

The paper is organized as follows.
In Sect.\,\ref{sec:setup}, we present 5-dimensional holographic models for heavy and light quarks and describe their thermodynamic properties. In Sect.\,\ref{sec:beta}, we describe the $\beta$-function $\beta(\alpha)$ for these models. In Sect.\,\ref{sec:concl}, we review our main results.

\newpage

\section{Holographic set up for light and heavy quarks models} \label{sec:setup}
\subsection{Backgrounds}

We consider Einstein-Maxwell-scalar (EMS) system with the action \cite{Li:2017tdz,Yang:2015aia}
\bea
S&=&\frac{1}{16\pi G_5}\int d^5x\sqrt{-g} \left[R-\frac{\ff_0(\varphi)}{4}F^2-\frac{1}{2}\partial_{\mu}\varphi\partial^{\mu}\varphi-\cV(\varphi)\right],\label{action}
\eea
where $G_5$ is the 5-dimensional Newtonian constant, $R$ is Ricci scalar, $F_{\mu\nu}=\partial_{\mu}A_{\nu}-\partial_{\nu}A_{\mu}$ is the electromagnetic tensor of the gauge Maxwell field $A_{\mu}$, $\varphi$ is the scalar (dilaton) field, $\ff_0(\varphi)$ is the gauge kinetic function describes the coupling between the Maxwell field and dilaton, $\cV(\varphi)$ is the dilaton potential.\\

Corresponding to (\ref{action}), equations of motion (EOMs) in general form can be obtained:
\bea\label{EOM1}
\nabla^2\varphi&=&\frac{\partial \cV}{\partial \varphi}+ \frac{F^2}{4}\frac{\partial \ff_0}{\partial \varphi} ~,\qquad \qquad \nabla_{\mu}\left[\ff_0(\varphi)F^{\mu\nu}\right]=0 ~,\\
\label{EOM2}
R_{\mu\nu}-\frac{1}{2}g_{\mu\nu}R&=&\frac{\ff_0(\varphi)}{2}\left(F_{\mu\rho}F^{\rho}_{\nu}-\frac{1}{4}g_{\mu\nu}F^2\right)+\frac{1}{2}\left[\partial_{\mu}\varphi\partial_{\nu}\varphi-\frac{1}{2}g_{\mu\nu}(\partial\varphi)^2-g_{\mu\nu}\cV(\varphi)\right].\nonumber\\
\eea
 To solve this system of equations, we consider the ansatz for the metric, scalar field and Maxwell field as \cite{Li:2017tdz,Yang:2015aia}
\bea
ds^2&=& B^2(z)\left[-g(z)dt^2+d\Vec{x}^2+\frac{dz^2}{g(z)}\right],\label{metric}\\
 \quad  \varphi &=&\varphi(z),\quad \label{warp-factor} A_{\mu}=\left(A_t(z)~,\Vec{0},0\right)~ , ~\, B(z)=\frac{e^{A(z)}}{z}\,,
\eea
where $B(z)$ is the warp factor, $g(z)$ is the blackening function, $\Vec{x}=(x_1,x_2,x_3)$, and  $A(z)$ is a scale  factor that has different functionality associated to the light and heavy quarks.
One can find the full explicit  EOMs in \cite{Li:2017tdz,Yang:2015aia,
  Arefeva:2018hyo, Arefeva:2022avn, Arefeva:2022bhx, Arefeva:2020vae,Arefeva:2020byn,Arefeva:2021mag,Arefeva:2023ter}. All functions  depend on the  holographic coordinate $z$, and after solving EOMs one can present all fields as functions of $\varphi$, i.e. $V(z)=\cV(\varphi(z))$,  
$V_\varphi(z)=\cV_\varphi(\varphi(z))$ and $f_0(z)=\ff_0(\varphi(z))$. \\

To solve the EOMs (\ref{EOM1})-(\ref{EOM2}), the usual boundary conditions (b.c) are used
\bea
  &&A_t(0) = \mu, \quad \, A_t(z_h) = 0, \label{eq:4.24} \\
 \quad  &&g(0) = 1, \qquad g(z_h) = 0, \,  \label{eq:4.25} \\
   &&\varphi(z,z_0)\Big|_{z=z_0}=0 \,.\qquad \qquad \label{eq:4.26}
\eea
Three different types of boundary conditions can be imposed on the  dilaton field by choosing different forms of $z_0$:
\bea \label{bc0}
z_0&=&0,\\
\label{bch}
z_0&=&z_h,\\ \label{bce}
z_0&=&\fz(z_h),
\eea
where $\fz (z_h)$ is a smooth function of $z_h$.
The dilaton field with zero boundary condition \eqref{bc0}  is denoted as $\varphi_0(z)$, i.e. $\varphi(z,0)=\varphi_0(z)$,
 the first boundary condition \eqref{bch} is shown as $\varphi_{z_h}(z)$ and the second boundary condition \eqref{bce} is considered as $\varphi_\fz(z)$, that for the light quarks model we will take \cite{AHSU}
\be\label{phi-fz-LQ}
z_0=\fz_{\,_{LQ}}(z_h)=10 \, e^{(-\frac{z_h}{4})}+0.1 \,.\ee
For the light quarks, one can choose a gauge kinetic function and  scale factor $A(z)$  in the form  \cite{Li:2017tdz}
\bea \label{wfLc}
f_0(z)=e^{-c \, z^2-A(z)},\quad
A(z)=-a\log(bz^2+1)~,
\eea
where $a = 4.046$, $b = 0.01613$   GeV${}^2$ and $c=0.227$   GeV${}^2$ are parameters that reproduce the  Regge trajectories from experiments  \cite{Li:2017tdz}.  
\\

For the heavy quarks, the gauge kinetic function and scale factor $A(z)$ are given by \cite{Yang:2015aia,Arefeva:2023jjh}
\bea \label{scaleHQ}
f_0(z)=e^{-\mathrm{s} \, z^2-A(z)},\qquad A(z)=-\frac{\mathrm{s}}{3}z^2- p\, z^4~,
\eea
where $\mathrm{s}= 1.16$   GeV${}^2$ and $p = 0.273$  GeV${}^4$ are parameters, that can be fitted with the experimental data \cite{Yang:2015aia}. 
 The second boundary condition for heavy quarks model was proposed as \cite{AHSU}
\bea \label{bceHQ}
z_0&=& \fz_{\,_{HQ}}(z_h) =e^{(-\frac{z_h}{4})} + 0.1\, .
\eea

 Under boundary conditions \eqref{eq:4.24}, \eqref{eq:4.25}, analytical solutions of the system of EOMs
(\ref{EOM1})-(\ref{EOM2}) for light and heavy quarks are
\bea
\label{phiprime} %,spsol,g,Vsol
\varphi(z)&=& \int_{z_0}^{z}\, dz
\sqrt{-6\Bigg(A''(z)-A'(z)^2+\frac{2}{z}A'(z) \Bigg)}\,,\label{spsol}\\
A_t(z)&=&\mu\frac{e^{cz^2}-e^{cz^2_h}}{1-e^{cz^2_h}}\,,
\\
g(z)&=&1-\frac{1}{\int_0^{z_h}y^3e^{-3A}dy}\Bigg[\int_0^zy^3 e^{-3A}dy-\frac{2c\mu^2}{(1-e^{cz_h^2})^2}\Bigg|\begin{matrix}
\int_0^{z_h}y^3e^{-3A}dy & \int_0^{z_h}y^3e^{-3A}e^{cy^2}dy \\
\int_{z_h}^z y^3e^{-3A}dy & \int_{z_h}^{z}y^3e^{-3A}e^{cy^2}dy
\end{matrix}\Bigg|\Bigg],\nonumber\\
\label{g}
\\\nn
\,
\\
V(z)&=&-3z^2g e^{-2A}\Bigg[A''+3A'^2+\Bigg(\frac{3g'}{2g}-\frac{6}{z}\Bigg)A'-\frac{1}{z}\Bigg(\frac{3g'}{2g}-\frac{4}{z}\Bigg)+\frac{g''}{6g}\Bigg]\, ,\label{Vsol}
\eea
where the symbol $'=d/dz$.  Note that for heavy quarks we need to replace the parameter $c$ with $\mathrm{s}$.

\subsection{Thermodynamics} \label{phaseLQ}

Using the metric \eqref{metric}, 
temperature and entropy can be written as:
\begin{gather}
  \begin{split}
    T &= \cfrac{|g'|}{4 \pi} \, \Bigl|_{z=z_h}
  \end{split}\, ,\qquad   s = \frac{{B}^{3}(z_h)}{4}. \label{eq:2.03}
\end{gather}
The entropy monotonically decreases with a horizon growth. To get a BH-BH (black hole - black hole, or Hawking-Page-like) phase transition line we need to calculate free energy as a function of temperature:
\begin{gather}
  F =  \int_{z_h}^{z_{h_2}} s \, T' dz. \label{eq:2.05}
\end{gather}
Consideration of non-negative values of the temperature implies that $z\leq z_{h_2}$. 

\begin{figure}[t!]
  \centering
\includegraphics[scale=0.39]{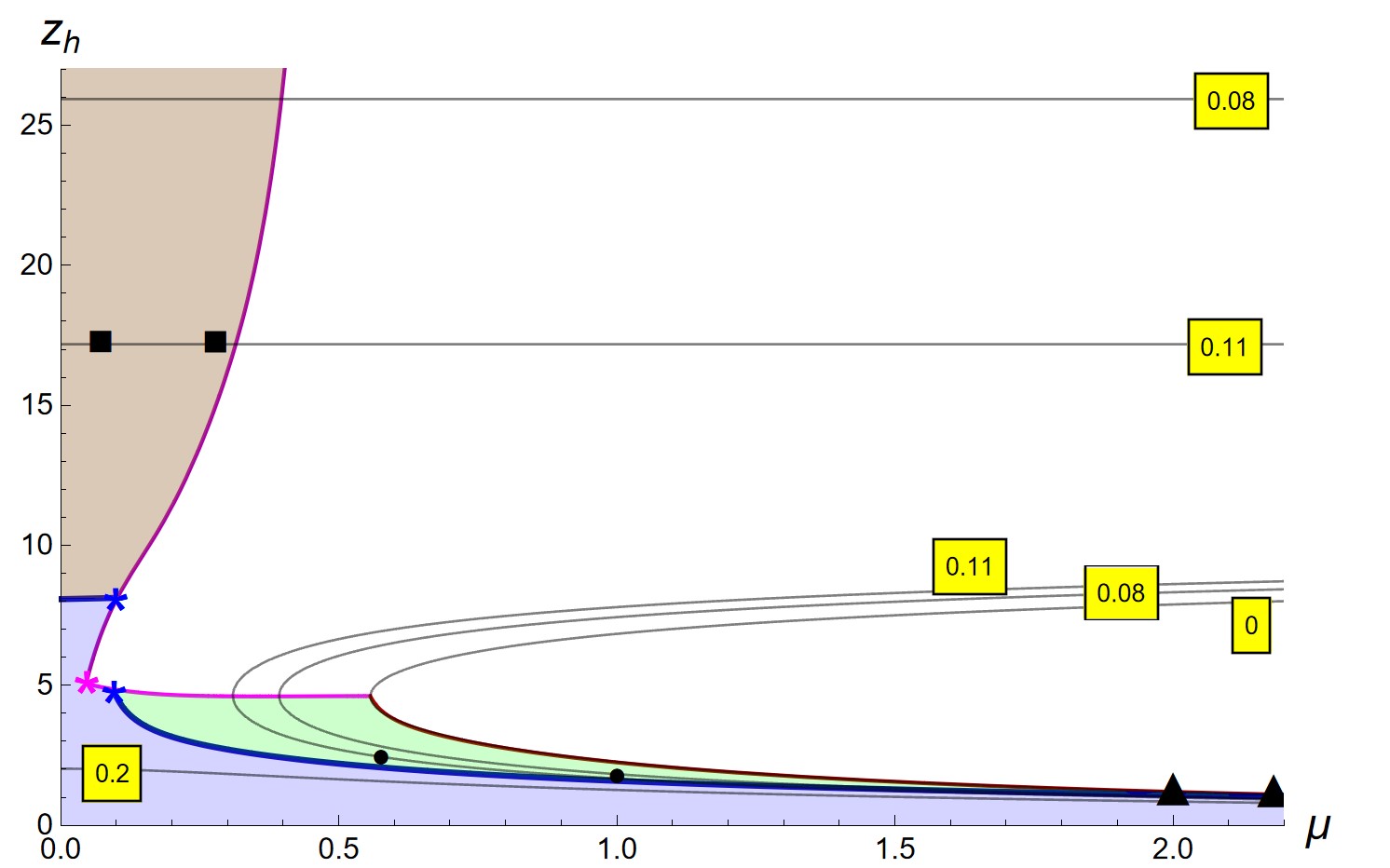}
\includegraphics[scale=0.4]{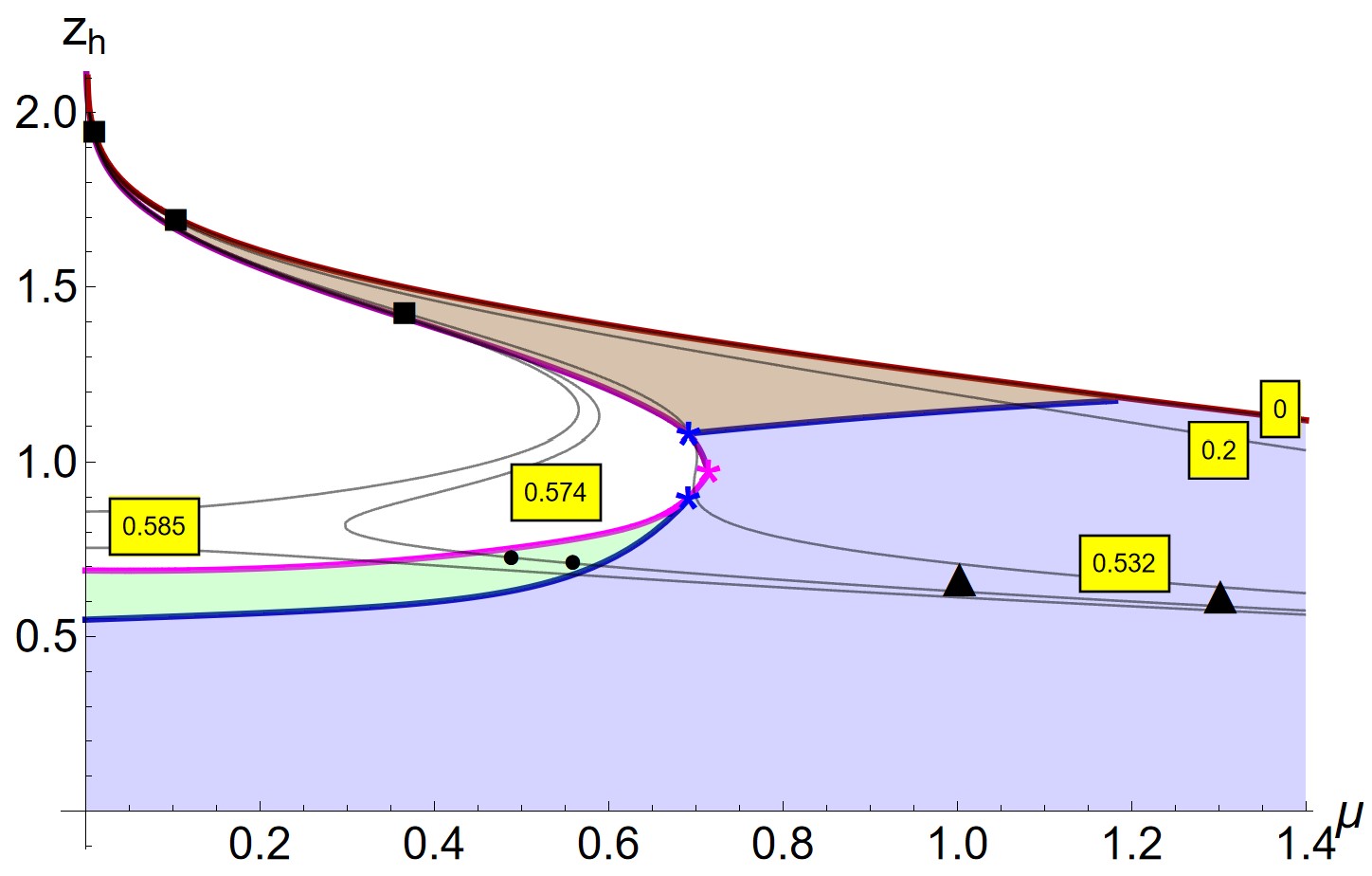}\\
A\hspace{15em}B
\caption{2D plots in $(\mu,z_h)$-plane for light quarks (A) and heavy quarks (B). Hadronic, quarkyonic and QGP phases are denoted by brown, green and blue, respectively.  Solid gray lines show the temperature indicated in rectangles. The intersection of the confinement/deconfinement and 1-st order phase transition lines is denoted by the blue stars. The magenta star indicates CEP; $[\mu]=[T]=[z_h]^{-1} =$ GeV.
%{\bf Math. file: LQ-T-zh-mu-phase-transitions.nb} 
% {\bf Math. file: HQ-T-zh-mu-phases-transitions-final.nb}.
}
\label{Fig:PhL2D}%{Fig:PhL3d}
\end{figure}

QCD phase diagram in $(\mu,T)$-plane describes the phase structure of quantum matter  in terms of thermodynamic parameters. The phase diagram for holographic models can be represented in $(\mu,z_h)$-plane, see Fig.\,\ref{Fig:PhL2D}A for light and Fig.\,\ref{Fig:PhL2D}B for heavy quarks \cite{AHSU}. Different phases, i.e. hadronic, quarkyonic, and QGP, are denoted by brown, green and blue  corresponding to squares, disks, and triangles, respectively. Solid gray lines show the constant temperatures indicated in rectangles, and the intersection of the confinement/deconfinement and 1-st order phase transition lines is denoted by the blue stars. The magenta star indicates critical end point (CEP). This figure describes the physical domains of the holographic model that we used to study the properties beta-function as a function of the running coupling.

\newpage

\section{Beta-Function} \label{sec:beta}

The dependence of running coupling  $\alpha$ on the energy scale of the physical system is described by beta-function, $\beta(\alpha)$.
Holographic $\beta$-function and running coupling are given by \cite{deBoer:1999tgo, Gursoy:2007cb,He:2010ye,Gursoy:2007er,Pirner:2009gr}
 \bea \label{beta-z}
\beta(\alpha)=3\,\alpha \,X, \, \, \qquad \qquad  \alpha(z)=e^{\varphi(z)},
 \eea
where $X$ is a new dynamical variable defined as \cite{Arefeva:2019qen,Arefeva:2018jyu,Arefeva:2020aan}
\be \label{X-z}
X(z)=\frac{\dot{\varphi} B}{3\dot{B}}\,,
\ee
$B(z)$ is specified in (\ref{warp-factor}) and $\varphi$ is obtained in (\ref{phiprime}).
Note that the function $X(z)$ does not depend on the choice of boundary conditions, although, 
the holographic $\beta$-function and coupling constant do depend.
We denote the dilaton field with the zero boundary condition at zero holographic coordinate as $\varphi=\varphi_0(z)$, i.e.
\be
\varphi_0(z)\Big|_{z=0}=0\,.
\ee
Considering the boundary condition at $z=z_0$, the dilaton field can be written as \cite{AHSU}
\be
\varphi_{z_0}(z)=\varphi_{0}(z)-\varphi_{0}(z_0).\ee
For the running coupling we have
\bea
&&\alpha_{0}(z)\to\alpha_{z_0}(z)=\alpha_0(z)\,\fG (z_0) \\ &&\mbox{where}
\quad\alpha_0(z)=e^{\varphi_0(z)},
\quad\fG (z_0)=e^{-\varphi_{0}(z_0)}
\eea
 and, then $\beta$-function can be written as
 \be\label{beta0}
\beta_0(z)\to   \beta_{z_0}(z)= \beta_0(z)\fG (z_0) \quad\mbox{with}\quad \beta_0(z)=3\alpha_0(z)X(z)\,. 
 \ee
While $\varphi_{0}(z)$ does not depend directly on the thermodynamic quantities such as $T$ and $\mu$ in our model, it is possible to include this dependence for running coupling representing $z_0$ in terms of $z_h$, i.e. $z_0=\fz(z_h)$, and get
\be\label{alpha_gen}
\alpha_{\fz}(z;T,\mu)=\alpha_0(z)\,\fG (T,\mu),\quad\mbox{where}\quad\fG (T,\mu)= e^{-\varphi_{0}(\fz(z_h))}\, ,
\ee
where $\fz(z_h)=z_h$ can be one choice, and another forms are given by exponential functions \eqref{phi-fz-LQ} and \eqref{bceHQ} for light and heavy quarks, respectively.\\

\subsection{$\beta$-function with different boundary conditions}

\begin{figure}[h!]
  \centering
\includegraphics[scale=0.58]{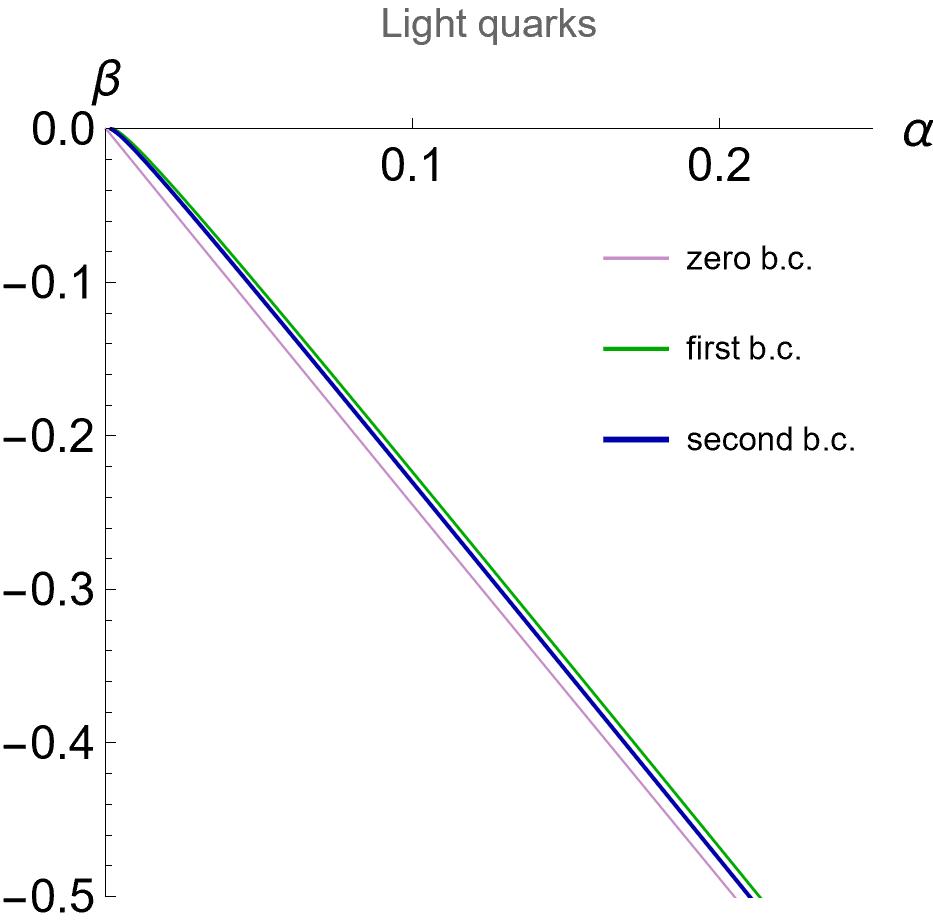}
\qquad 
\includegraphics[scale=0.56]{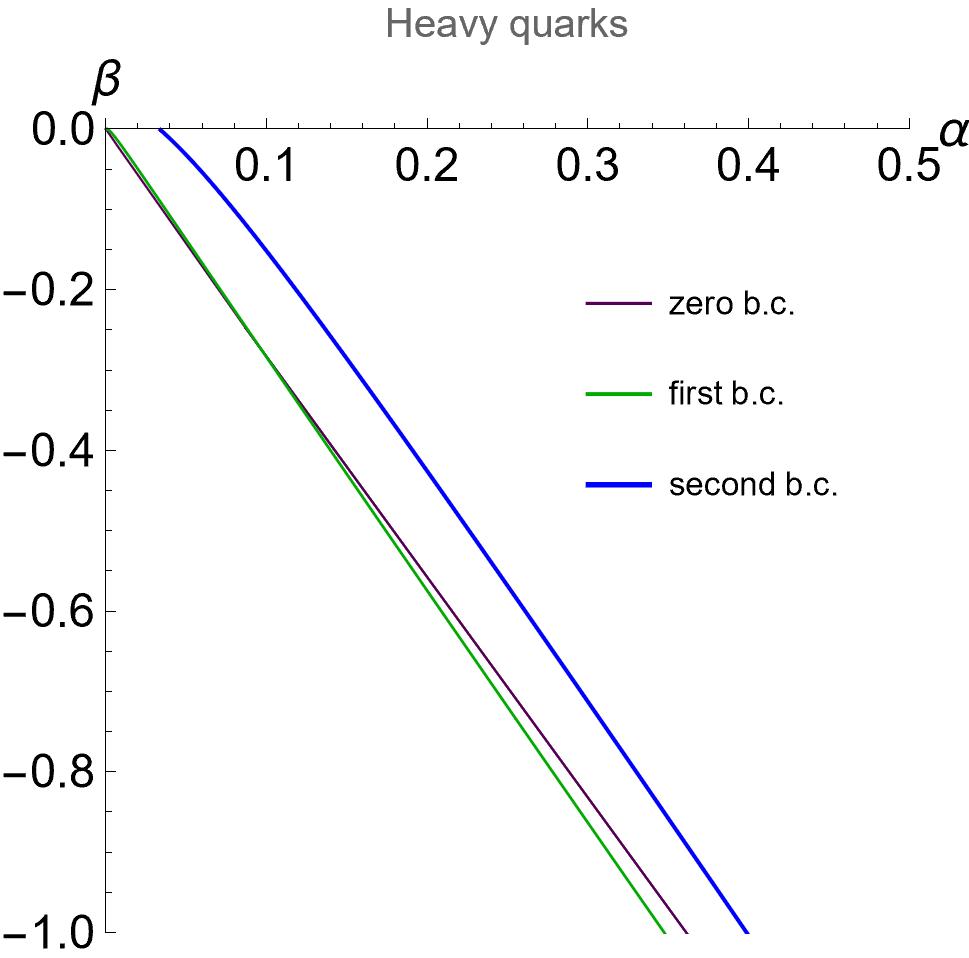}\\
A\hspace{225pt}B
\caption{Beta-function $\beta(\alpha)$ for light quarks model at $T=0.08$ and $\mu=0.43$ (A) and heavy quarks model  at $T=0.532$ and $\mu=0.64$ (B) with different boundary conditions; $[\mu]=[T]=$ GeV.}
%{\bf Math. file:  2D-beta.nb}
\label{beta-bc1}
\end{figure}

The beta-function $\beta(\alpha)$ for light quarks model at $T=0.08$, $\mu=0.43$ and for heavy quarks model at $T=0.532$, $\mu=0.64$ with  three different types of boundary conditions, i.e. \eqref{bc0}, \eqref{bch} and \eqref{bce}, are depicted in Fig.\,\ref{beta-bc1}.
The behavior of $\beta(\alpha)$ does not crucially depend on the boundary conditions both for light  and heavy quarks. For this reason, in this paper we chose the first boundary condition \eqref{bce} to investigate the properties of the beta-function $\beta(\alpha)$.

 \newpage

\subsection{$\beta$-function as a function of running coupling }

\begin{figure}[h!]
  \centering
\includegraphics[scale=0.2]{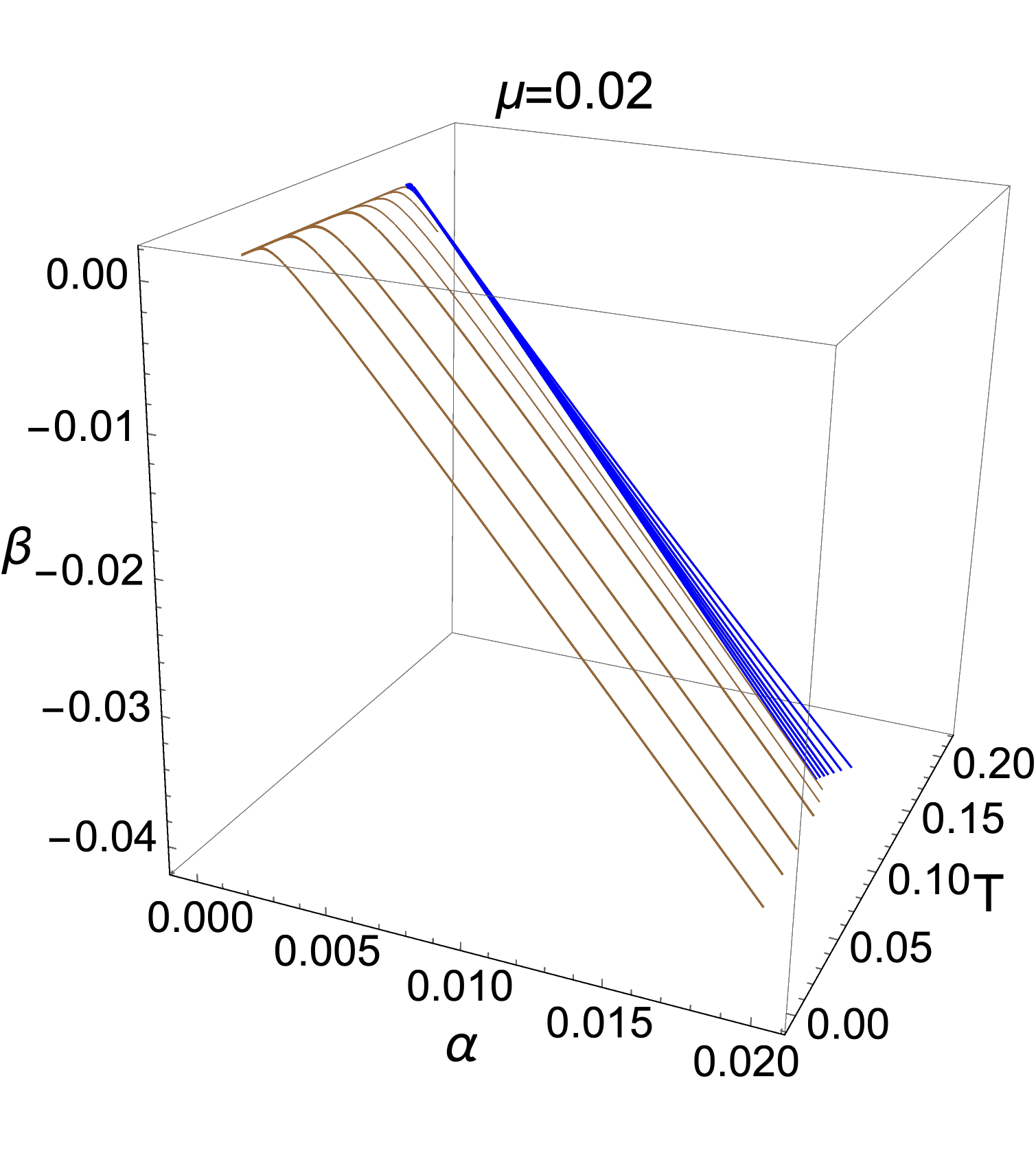} \,
 \includegraphics[scale=0.27]{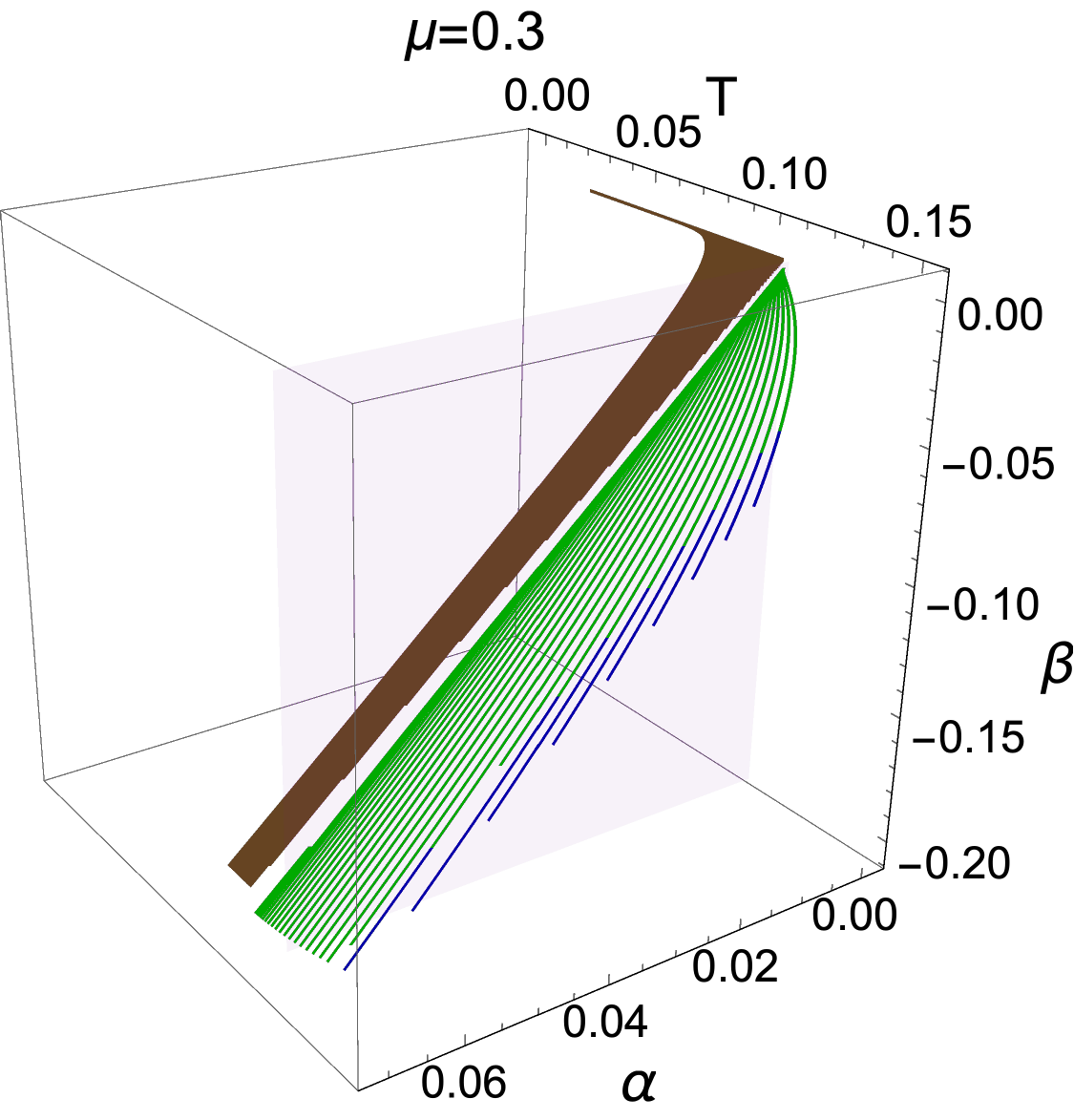} \quad
\includegraphics[scale=0.23]{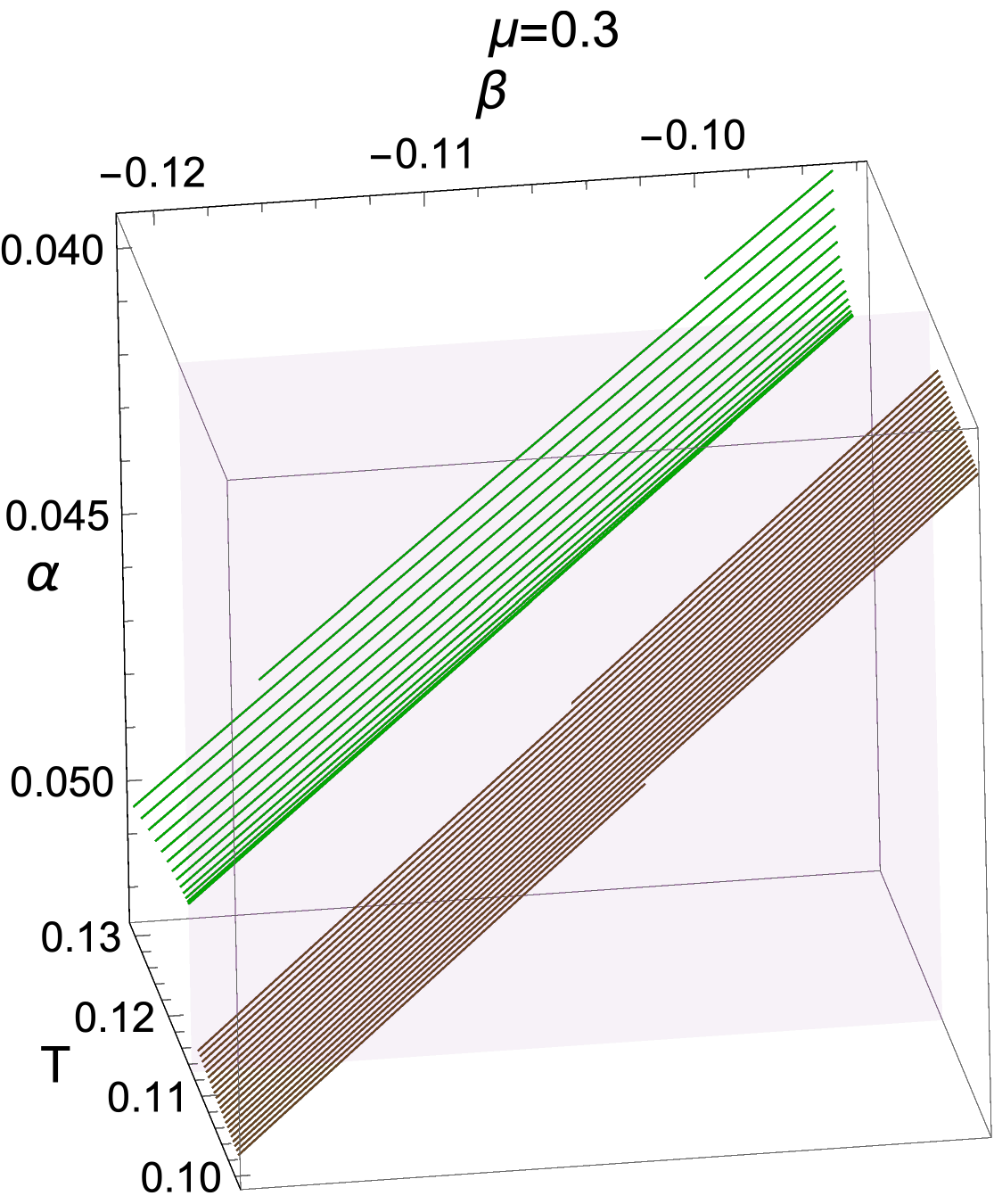}\\
\qquad\qquad\qquad \qquad \qquad \qquad
\includegraphics[scale=0.43]{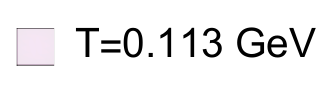}\\
A\hspace{12em}B\hspace{12em}C
\caption{3D-plot of beta-function $\beta(\alpha;\mu,T)$ for light quarks at fixed $\mu=0.02$ (A) $\mu=0.3$ (B) and its zoom at the 1-st order phase transition temperature $T=0.113$ shown by the light red plane (C). Hadronic, quarkyonic and QGP phases are denoted by brown, green and blue, respectively; $[\mu]=[T]=$ GeV. 
%{\bf Math. files: LQ-PT-coupling-beta-alpha-mu002-final.nb, LQ-PT-coupling-beta-alpha-mu03-final.nb}
} \label{Fig:beta(alpha)-T-mu}
\end{figure}

\begin{figure}[h!]
  \centering
\includegraphics[scale=0.55]{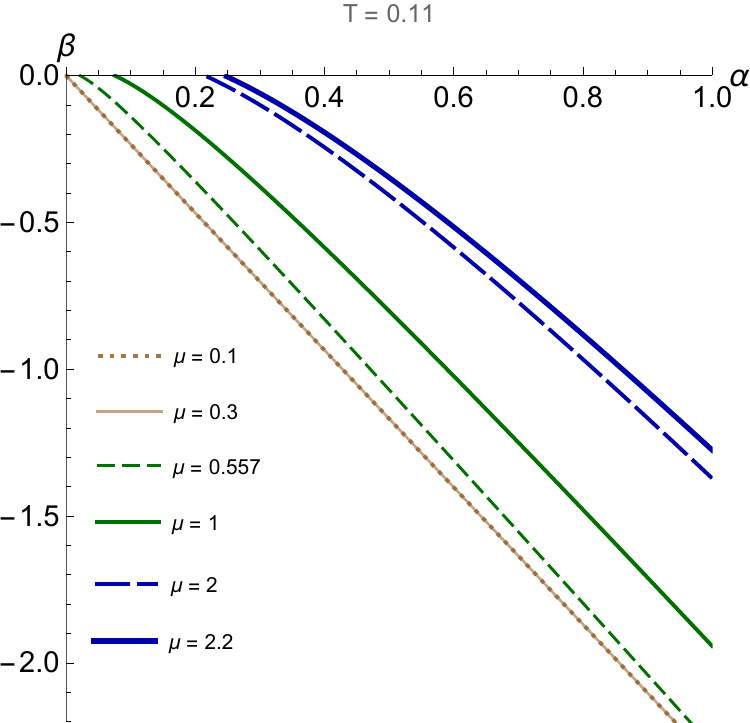}\qquad \includegraphics[scale=0.55]{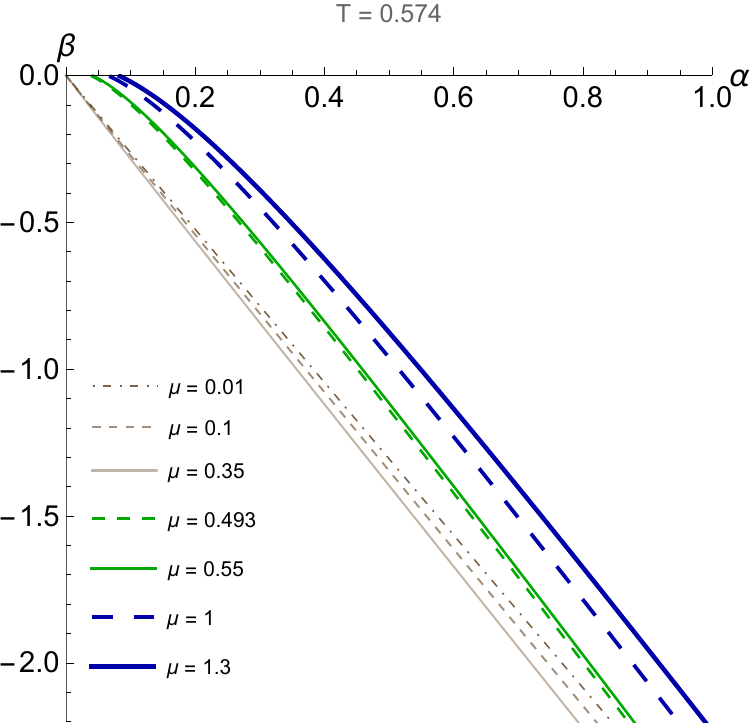}\\
A\hspace{17em}B
\caption{2D-plot of beta-function $\beta=\beta(\alpha)$ for light quarks at $T=0.11$ (A)  and for heavy quarks at $T=0.574$ (B) with different $\mu$; $[\mu]=[T]=$ GeV. 
%{\bf Math.file: beta-LQ-2D-TMP.nb, beta-HQ-2D-TMP.nb}
}
\label{Fig:LQ-beta(alpha)-2D-2}
\end{figure}

The 3D-plot of beta-function $\beta(\alpha;\mu,T)$ for light quarks at fixed $\mu=0.02$ (A), $\mu=0.3$ (B) and its zoom at the 1-st order phase transition temperature $T=0.113$ (C) is plotted in Fig.\,\ref{Fig:beta(alpha)-T-mu}.  Hadronic, quarkyonic and QGP phases are denoted by brown, green and blue,  respectively. The constant temperature $T=0.113$ is shown by light red plane. The Fig.\,\ref{Fig:beta(alpha)-T-mu}A shows the crossover region where the 1-st order phase transition does not occur, while confinement/deconfinement phase transition occurs and $\beta(\alpha;\mu,T)$ has no jump from hadronic to QGP phases. Although, in the Fig.\,\ref{Fig:beta(alpha)-T-mu}B from quarkyonic to QGP phases there is a continuous phase transition without any jump while from hadronic to quarkyonic phases the  $\beta(\alpha;\mu,T)$ has a jump.
These jumps in the values of the $\beta$-function during the 1-st order phase transition on different scales (parameterized by the holographic coordinate $z$) are shown in the Fig.\,\ref{Fig:beta(alpha)-T-mu}C. The Fig.\,\ref{Fig:beta(alpha)-T-mu} shows that the dependence of $\beta$-function on $\alpha$ is mainly linear with a slope depending on the parameters $T$ and $\mu$. To make this behavior more clear, the $\beta$-function as a function of $\alpha$ for light quarks at  different values of  $\mu$ is plotted, see 
Fig.\,\ref{Fig:LQ-beta(alpha)-2D-2}A.

\begin{figure}[h!]
  \centering
\includegraphics[scale=0.25]{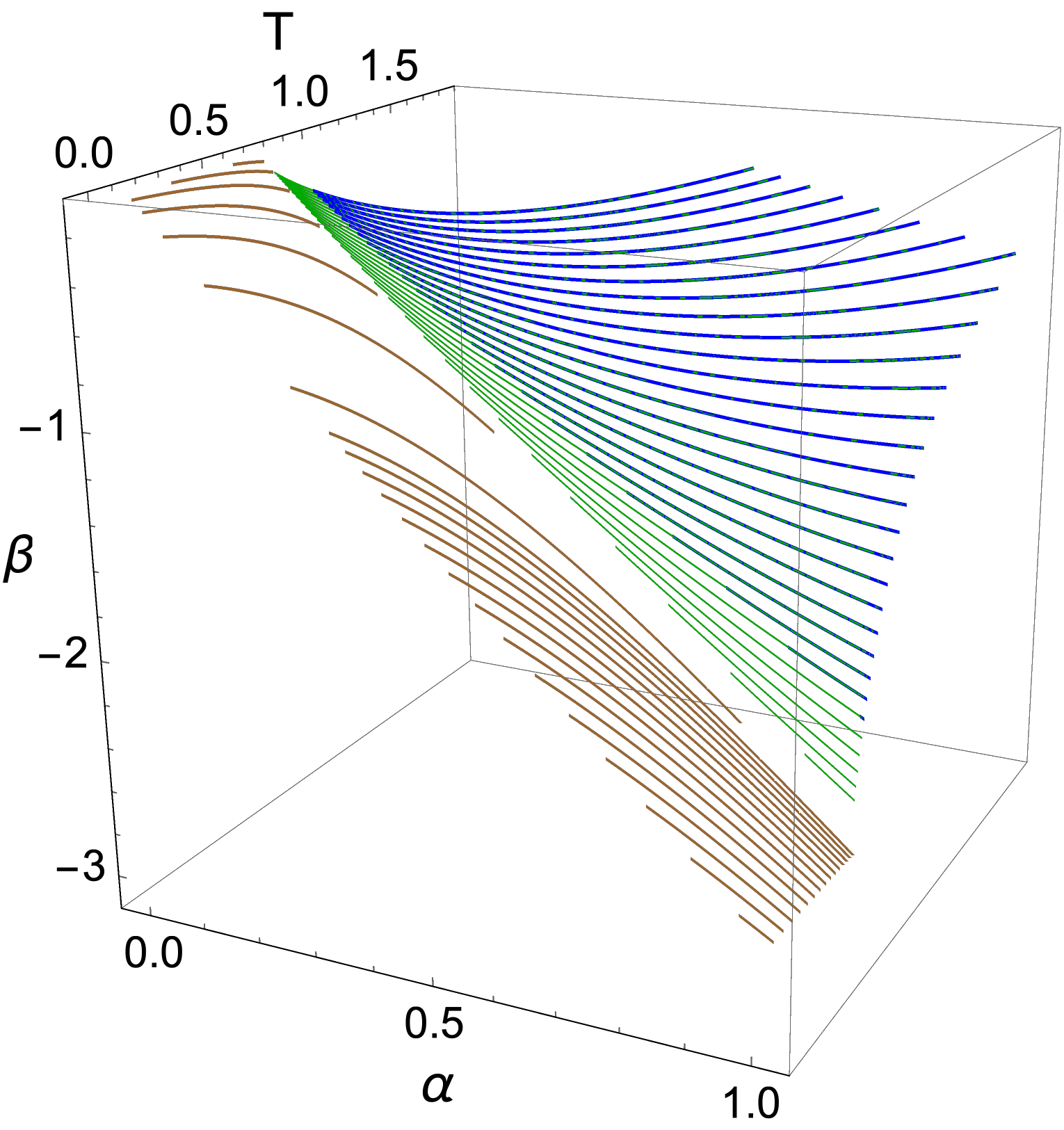}
\qquad \includegraphics[scale=0.27]{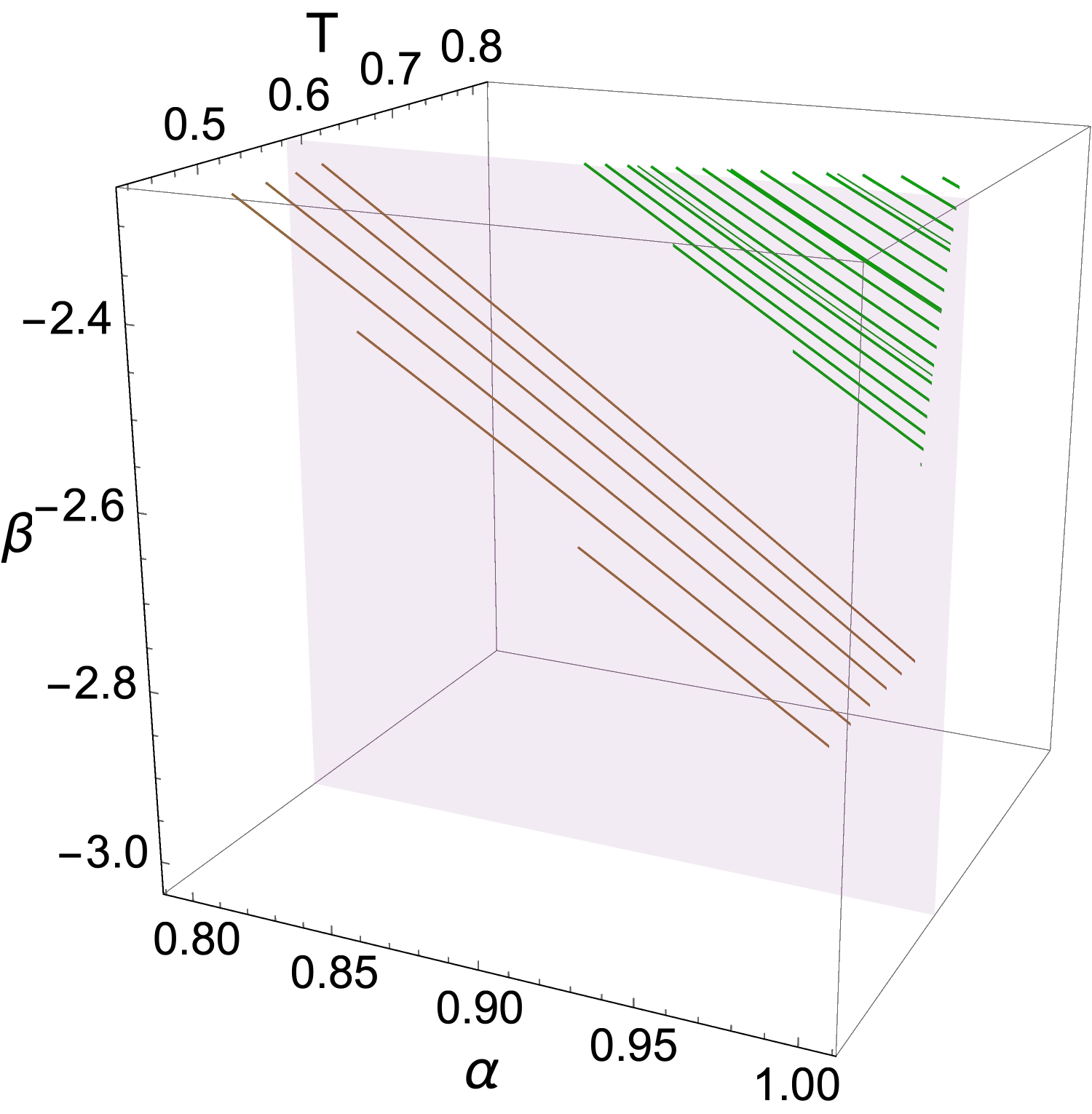}\\
\qquad \qquad \qquad \qquad \qquad 
\includegraphics[scale=0.36]{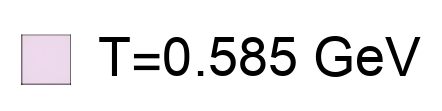}
\\
A\hspace{150pt}B\\
\caption{3D-plot of beta-function $\beta(\alpha;\mu,T)$ for heavy quarks at fixed $\mu=0.3$ (A) and its zoom at the 1-st order phase transition temperature $T=0.585$ (B). Hadronic, quarkyonic and QGP phases are denoted by brown, green and blue, respectively. The plane with  $T=0.585$, is shown by light red; $[\mu]=[T]=$ GeV.
%{\bf Math. files: HQ-LQ-beta(alpha)-mu03-08.nb}
}
 \label{Fig:beta(alpha)-T-mu-HQ}
\end{figure}

To investigate the jumps in the values of the $\beta$-function for heavy quarks model during the 1-st order phase transition on different scales parameterized by the holographic coordinate $z$, we represent the beta-function $\beta(\alpha;\mu,T)$  at fixed $\mu=0.3$ (A) and its zoom at the 1-st order phase transition temperature $T=0.585$ (B) in Fig.\,\ref{Fig:beta(alpha)-T-mu-HQ}. Hadronic, quarkyonic and QGP phases are denoted by brown, green and blue,  respectively. The plane with  $T=0.585$, is shown by light red.  In Fig.\,\ref{Fig:beta(alpha)-T-mu-HQ}A 
there is confinement/deconfinement phase transition occurs without any jump between quarkyonic and QGP phases and there is a jump between hadronic and quarkyonic phases at the 1-st order phase transition. To see the jump in detail,  
Fig.\,\ref{Fig:beta(alpha)-T-mu-HQ}B is depicted where the light red plane shows the $T=0.585$ of the 1-st order phase transition. 
In Fig.\,\ref{Fig:beta(alpha)-T-mu-HQ}, the dependence of $\beta$-function on $\alpha$ is mostly linear with a slope depending on the parameters $T$ and $\mu$. To have more clarification, we plotted $\beta$-function as a function of $\alpha$ for different values of $\mu$,  see Fig.\,\ref{Fig:LQ-beta(alpha)-2D-2}B.

It is very important to note that our results for light and heavy quarks show that at the 1-st order phase transition $\beta$-function experiences the jumps depending on the parameters of the theory, i.e $T$ and $\mu$. In fact, the $\beta$-function can feel the 1-st order phase transition.

In Fig.\,\ref{Fig:LQ-beta(alpha)-2D-2} 2D-plot of beta-function $\beta=
\beta(\alpha)$ for light quarks at $T=0.11$ (A)  and for heavy quarks at $T=0.574$ (B) with different $\mu$ is shown.
The Fig.\,\ref{Fig:LQ-beta(alpha)-2D-2}A describes the light quarks $\beta(\alpha)$ at $\mu=0.1,0.3$, $\mu=0.557,1$ and $\mu=2,2.2$ corresponding to hadronic, quarkyonic and QGP phases, respectively.
In the hadronic phase of the Fig.\,\ref{Fig:LQ-beta(alpha)-2D-2}A, the $\beta(\alpha)$ at fixed $T$ 
does not depend on different $\mu$. This result is consistent with \cite{AHSU}. 
The Fig.\,\ref{Fig:LQ-beta(alpha)-2D-2}B describes the heavy quarks $\beta(\alpha)$ at $\mu=0.01,0.1,0.35$, $\mu=0.493,0.55$ and $\mu=1,1.3$ corresponding to hadronic, quarkyonic and QGP phases, respectively.
In the hadronic phase of the heavy quarks, the $\beta(\alpha)$ at fixed $T$ depends on different values of $\mu$. This result is consistent with \cite{AHSU}. 
For all cases in Fig.\,\ref{Fig:LQ-beta(alpha)-2D-2}, the $\beta$-function is negative and  decreases with increasing $\alpha$. 
For both light and heavy quarks model the beta-function increases from hadronic to quarkyonic and then to QGP phases. % \MU{with increasing of $T$ \it (note: we can't state that $\beta(\alpha)$ for HQ in hadronic phase demostrates increasing behavior when $\mu$ increases)}.
The beta-function, $\beta$, has the linear dependence on $\alpha$ in all cases, although in the Quarkyonic and QGP phases of light and heavy quarks for small values of holographic coordinate $z$ there are deviations from linear behavior. It can be described by noting that  $\beta(\alpha)$ \eqref{beta-z} depends on the $X=X(z)$ that qualitatively at small $z$ the $X(z)$ changes crucially and at large $z$ is approximately constant.

To obtain the physical results for $\beta$-function for light and heavy quarks model as a function of different parameters of the theory, we must respect the physical domains of different phases that has already been obtained in Fig.\,\ref{Fig:PhL2D}.

$$\,$$\newpage
%$$\,$$\newpage

\newpage

\subsection{Comparing holographic $\beta$-function and perturbation results}\label{Pert-Holo}

\begin{figure}[h!]
  \centering
\includegraphics[scale=0.5]{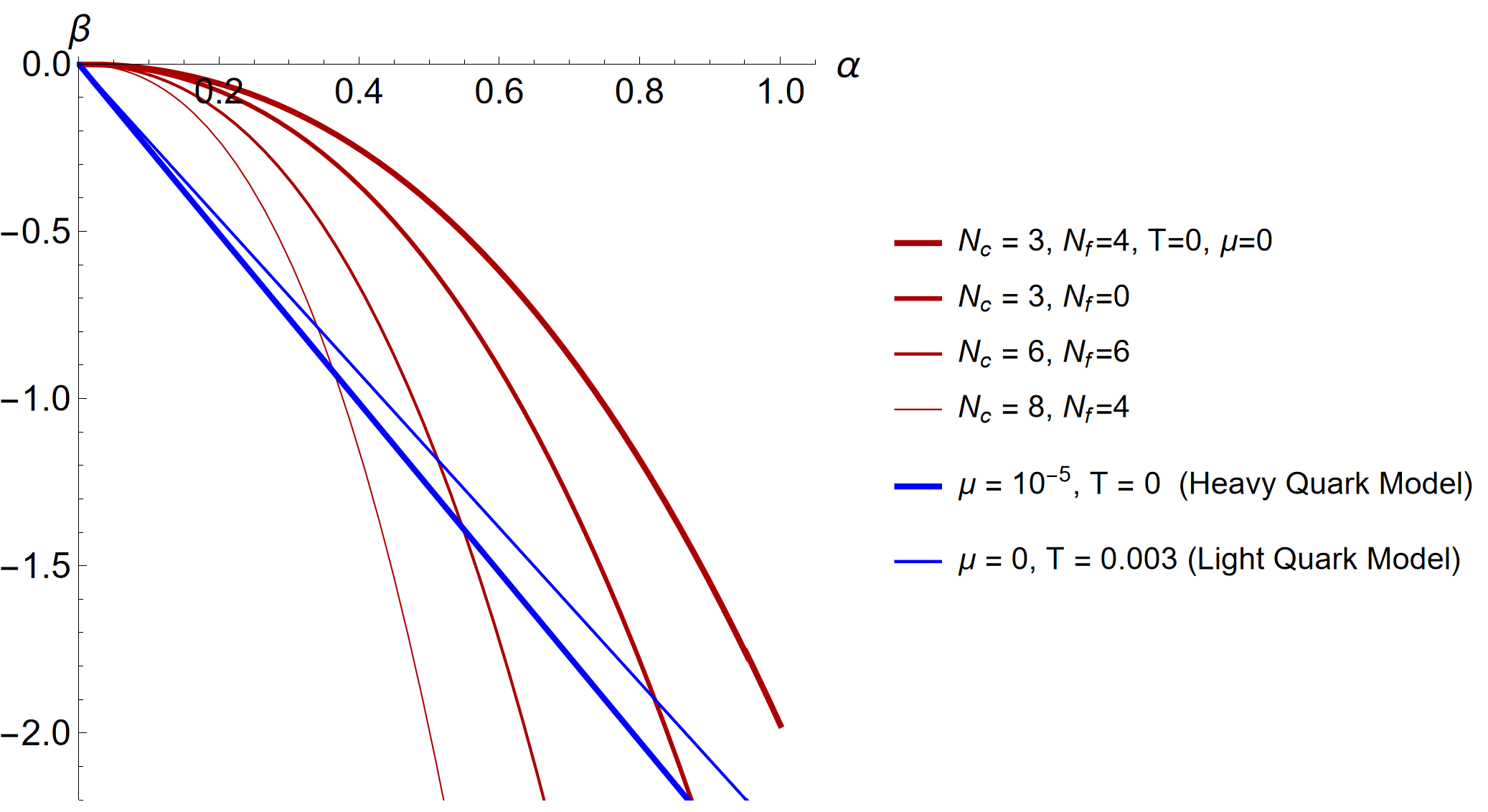}
\caption{Beta-function $\beta(\alpha)$ for QCD at 2-loop level for $T=0$, $\mu=0$ at different $N_c$ and $N_f$ in red lines, and holographic $\beta$-function for light quarks at $\mu=0$, $T=0.003$ (light blue) and heavy quarks $\mu=10^{-5}$, $T=0$ (dark blue); $[\mu]=[T]=$ GeV. %{\bf Math. files: beta-alpha-lattice-LQ.nb}
} \label{Fig:beta-lattice-LQ1}
\end{figure}

The QCD $\beta$-function at 2-loop level has the following form \cite{vanRitbergen:1997va,He:2010ye}\be \label{beta-perturb}
\beta(\alpha)=-b_0 \, \alpha^{2} - b_1 \, \alpha^3 \, ,
\ee
where $$b_0=\frac{1}{2\pi}\left(\frac{11}{3}N_c - \frac{2}{3} N_f \right)$$ 
and 
$$b_1=\frac{1}{8\pi^2}\left( \frac{34}{3}N_c^2-\left(\frac{13}{3}N_c-\frac{1}{N_f}\right)\right)\, ,$$ 
where $N_c$ is number of colors and $N_f$ is number of flavors. The $\beta$-function $\beta(\alpha;\mu,T)$ for QCD at the 2-loop level for $T=0$, $\mu=0$, at different $N_c$ and $N_f$ is shown in red lines. The holographic $\beta$-function for light quarks at $\mu=0$, $T=0.003$ (light blue) and for heavy quarks at $\mu=10^{-5}$, $T=0$ (dark blue) are plotted in Fig.\,\ref{Fig:beta-lattice-LQ1}. The $\beta$-function produced via holography is negative and monotonically decreases with increasing coupling constant $\alpha$, agreeing reasonably well with the QCD $\beta$-function obtained using perturbative calculations, shown by the red lines.

\newpage

\section{Conclusion} \label{sec:concl}

In this paper, we considered the $\beta$-function in isotropic holographic models supported by the Einstein-dilaton-Maxwell action for heavy and light quarks. We investigated the $\beta$-function's dependence on the running coupling for various chemical potentials and temperatures. At 1-st-order phase transitions, $\beta$-functions exhibit discontinuities depending on temperature and chemical potential.
\\

To present the $\beta$-function as a function of $\alpha$, we imposed a boundary condition on the dilaton field. In \cite{AHSU}, three different boundary conditions are introduced, and since there is no significant difference in the behavior of $\beta(\alpha)$, we chose one of them.
\\

For both light and heavy quarks:
\begin{itemize}
  \item The $\beta$-function exhibits a jump at the 1-st order phase transition line, while there is no jump in the $\beta$-function at the confinement/deconfinement phase transition.
  \item $\beta(\alpha)$ is negative and continuously decreases with increasing $\alpha$, which is generally consistent with perturbative results.
  \item $\beta(\alpha)$ increases from the hadronic phase to the quarkyonic phase and from the quarkyonic phase to the QGP phase.
  \item The dependence of $\beta$ on $\alpha$ is primarily linear with a slope depending on the parameters $T$ and $\mu$, although deviations from linear behavior occur in the quarkyonic and QGP phases for small values of the holographic coordinate $z$.
\end{itemize}

Furthermore, $\beta(\alpha)$ in the hadronic phase exhibits a difference between light and heavy quarks: it assumes the same values for different $\mu$ in the light quarks model, whereas it varies with $\mu$ for heavy quarks. This observation is consistent with the running coupling results reported in \cite{AHSU}.
\\

Moreover, the choice of the warp factor $B(z)$ can significantly influence the qualitative behavior of $\beta(\alpha)$. By modifying the holographic model, one can potentially align the $\beta$-function even more closely with perturbation theory predictions, which we plan to explore in future work. Additionally, spatial anisotropy is known to affect the QCD phase transition temperature \cite{Arefeva:2018hyo,Arefeva:2018cli, Arefeva:2020uec, Chen:2024jet}. Another type of anisotropy arises from magnetic fields and their impact on the QCD phase diagram is also documented \cite{Gursoy:2017wzz,Bohra:2019ebj,Bohra:2020qom,Arefeva:2020vae,Dudal:2021jav,Jain:2022hxl,Arefeva:2022avn,Arefeva:2024xmg}. This form of anisotropy also modifies the $\beta$-function, which we intend to investigate in a separate paper.

\section{Acknowledgements}

The work of I.A., P.S. and M.U. is supported by RNF grant 20-12–00200. 
The work of A.H. was performed at the Steklov
International Mathematical Center  and  supported by the Ministry of
Science and Higher Education of the Russian Federation (Agreement
No.075-15-2022-265).
%\IA{\bf division between us}

%%%%%%%%%%%%%%%%%%%%%%%%%%%%%%%%%
\, \\


\newpage
\begin{thebibliography}{99}

\bibitem{AHSU}
I.~Y.~Aref'eva, A.~Hajilou, P.~Slepov and M.~Usova,
``Running Coupling for Holographic QCD with Heavy and Light Quarks: Isotropic case'', the first part of [arXiv:2402.14512 [hep-th]].

\bibitem{Maldacena:1997re}
J.~M.~Maldacena,
``The Large N limit of superconformal field theories and supergravity,''
Adv. Theor. Math. Phys. \textbf{2}, 231-252 (1998)
%doi:10.4310/ATMP.1998.v2.n2.a1
[arXiv:hep-th/9711200 [hep-th]].
\bibitem{Casalderrey-Solana:2011dxg}
J.~Casalderrey-Solana, H.~Liu, D.~Mateos, K.~Rajagopal and
  U.~A.~Wiedemann,
  ``Gauge/String Duality, Hot QCD and Heavy Ion Collisions'',
  (Cambridge University Press, Cambridge,
  UK, 2014), 
  % ISBN 978-1-139-13674-7
  % doi:10.1017/CBO9781139136747
  [arXiv:1101.0618 [hep-th]]

\bibitem{Arefeva:2014kyw}
 I.~Y.~Aref'eva,
  ``Holographic approach to quark-gluon plasma in heavy ion collisions'',
  Phys. Usp. \textbf{57}, 527-555 (2014)
  % doi:10.3367/UFNe.0184.201406a.0569

\bibitem{Boonstra:1998mp}
H.~J.~Boonstra, K.~Skenderis and P.~K.~Townsend,
``The domain wall / QFT correspondence,''
JHEP \textbf{01}, 003 (1999)
%doi:10.1088/1126-6708/1999/01/003
[arXiv:hep-th/9807137 [hep-th]].

\bibitem{Heemskerk:2010hk}
I.~Heemskerk and J.~Polchinski,
``Holographic and Wilsonian Renormalization Groups,''
JHEP {\bf 1106} (2011) 031
  %doi:10.1007/JHEP06(2011)031
  [arXiv:1010.1264 [hep-th]].

\bibitem{Kiritsis:2014kua}
E.~Kiritsis, W.~Li and F.~Nitti,
``Holographic RG flow and the Quantum Effective Action,''
Fortsch. Phys. \textbf{62}, 389-454 (2014)
%doi:10.1002/prop.201400007
[arXiv:1401.0888 [hep-th]].

\bibitem{Gursoy:2018umf}
U.~G\"ursoy, E.~Kiritsis, F.~Nitti and L.~Silva Pimenta,
``Exotic holographic RG flows at finite temperature,''
JHEP \textbf{10}, 173 (2018)
%doi:10.1007/JHEP10(2018)173
[arXiv:1805.01769 [hep-th]].

\bibitem{BogSchirkov}
N.N. Bogoliubov and  D.V. Shirkov,   The Theory of Quantized Fields. GTTL, Moscow (1957); English trans.: New York, NY: Interscience (1959).
%\cite{Wilson:1973jj}
\bibitem{Wilson:1973jj}
K.~G.~Wilson and J.~B.~Kogut,
``The Renormalization group and the epsilon expansion,''
Phys. Rept. \textbf{12}, 75-199 (1974).
%doi:10.1016/0370-1573(74)90023-4

\bibitem{Callan:1970yg}%{Symanzik:1970rt}
C.~G.~Callan, Jr.,
``Broken scale invariance in scalar field theory,''
Phys. Rev. D \textbf{2}, 1541-1547 (1970)
%doi:10.1103/PhysRevD.2.1541
%1266 citations counted in INSPIRE as of 
%\cite{Symanzik:1970rt}
\bibitem{Symanzik:1970rt}
K.~Symanzik,
``Small distance behavior in field theory and power counting,''
Commun. Math. Phys. \textbf{18}, 227-246 (1970)
%doi:10.1007/BF01649434
%1259 citations counted in INSPIRE as of 18 Nov 2023

\bibitem{Brown:1990ev}
  F.~R.~Brown, F.~P.~Butler, H.~Chen, N.~H.~Christ, Z.~h.~Dong,
  W.~Schaffer, L.~I.~Unger and A.~Vaccarino,
  ``On the existence of a phase transition for QCD with three light
  quarks'',
  Phys. Rev. Lett. \textbf{65}, 2491-2494 (1990)
  % doi:10.1103/PhysRevLett.65.2491

\bibitem{Philipsen:2016hkv}
O.~Philipsen and C.~Pinke,
``The $N_f=2$ QCD chiral phase transition with Wilson fermions at zero and imaginary chemical potential,''
Phys. Rev. D \textbf{93}, no.11, 114507 (2016)
%doi:10.1103/PhysRevD.93.114507
[arXiv:1602.06129 [hep-lat]]. 

\bibitem{Fu:2019hdw}
W.~j.~Fu, J.~M.~Pawlowski and F.~Rennecke,
``QCD phase structure at finite temperature and density,''
Phys. Rev. D \textbf{101}, no.5, 054032 (2020)
%doi:10.1103/PhysRevD.101.054032
[arXiv:1909.02991 [hep-ph]].



\bibitem{Li:2017tdz}
M.~W.~Li, Y.~Yang and P.~H.~Yuan,
``Approaching Confinement Structure for Light Quarks in a Holographic Soft Wall QCD Model,''
Phys. Rev. D \textbf{96}, no.6, 066013 (2017)
%doi:10.1103/PhysRevD.96.066013
[arXiv:1703.09184 [hep-th]].

\bibitem{Yang:2015aia}
Y.~Yang and P.~H.~Yuan,
``Confinement-deconfinement phase transition for heavy quarks in a soft wall holographic QCD model,''
JHEP \textbf{12}, 161 (2015)
%doi:10.1007/JHEP12(2015)161
[arXiv:1506.05930 [hep-th]].

\bibitem{Arefeva:2018hyo}
I.~Aref'eva and K.~Rannu,
``Holographic Anisotropic Background with Confinement-Deconfinement Phase Transition,''
JHEP \textbf{05}, 206 (2018)
%doi:10.1007/JHEP05(2018)206
[arXiv:1802.05652 [hep-th]].

\bibitem{Arefeva:2022bhx}
  I.~Y.~Aref\textquoteright{}eva, K.~A.~Rannu and P.~S.~Slepov,
  ``Anisotropic solution of the holographic model of light quarks with
  an external magnetic field'',
  Theor. Math. Phys. \textbf{210}, no.3, 363-367 (2022).
  % doi:10.1134/S0040577922030060

  \bibitem{Arefeva:2020byn}
I.~Y.~Aref'eva, K.~Rannu and P.~Slepov,
``Holographic anisotropic model for light quarks with confinement-deconfinement phase transition,''
JHEP \textbf{06}, 090 (2021)
%doi:10.1007/JHEP06(2021)090
[arXiv:2009.05562 [hep-th]].

\bibitem{Arefeva:2021mag}
  I.~Y.~Aref'eva, K.~Rannu and P.~S.~Slepov,
  ``Anisotropic solutions for a holographic heavy-quark model with an
  external magnetic field'',
  Teor. Mat. Fiz. \textbf{207}, no.1, 44-57 (2021)
  % doi:10.1134/S0040577921040036

\bibitem{Andreev:2006ct}
O.~Andreev and V.~I.~Zakharov,
``Heavy-quark potentials and AdS/QCD,''
Phys. Rev. D \textbf{74}, 025023 (2006)
%doi:10.1103/PhysRevD.74.025023
[arXiv:hep-ph/0604204 [hep-ph]].

\bibitem{Arefeva:2016rob}
I.~Aref'eva,
``Holography for Heavy Ions Collisions at LHC and NICA,''
EPJ Web Conf. \textbf{164}, 01014 (2017)
%doi:10.1051/epjconf/201716401014
[arXiv:1612.08928 [hep-th]].

  \bibitem{Gursoy:2007cb}
U.~Gursoy and E.~Kiritsis,
``Exploring improved holographic theories for QCD: Part I,''
JHEP \textbf{02}, 032 (2008)
%doi:10.1088/1126-6708/2008/02/032
[arXiv:0707.1324 [hep-th]].

\bibitem{Gursoy:2007er}
  U.~Gursoy, E.~Kiritsis and F.~Nitti,
  ``Exploring improved holographic theories for QCD: Part II'',
  JHEP \textbf{02}, 019 (2008)
  % doi:10.1088/1126-6708/2008/02/019
  [arXiv:0707.1349 [hep-th]].

\bibitem{Arefeva:2019qen}
I.~Y.~Aref'eva,
``Holographic Renormalization Group Flows,''
Theor. Math. Phys. \textbf{200}, no.3, 1313-1323 (2019)
%doi:10.1134/S0040577919090058

\bibitem{Arefeva:2020aan}
I.~Y.~Aref'eva and K.~Rannu,
``Holographic Renormalization Group Flow in Anisotropic Matter,''
Theor. Math. Phys. \textbf{202}, no.2, 272-283 (2020).
%doi:10.1134/S0040577920020105

\bibitem{Arefeva:2018jyu}
I.~Y.~Aref'eva, A.~A.~Golubtsova and G.~Policastro,
``Exact holographic RG flows and the A$_{1}$ \texttimes{} A$_{1}$ Toda chain,''
JHEP \textbf{05}, 117 (2019)
%doi:10.1007/JHEP05(2019)117
[arXiv:1803.06764 [hep-th]].

\bibitem{Golubtsova:2022hfk}
A.~A.~Golubtsova and M.~K.~Usova,
``Stability analysis of holographic RG flows in 3d supergravity,''
Eur. Phys. J. Plus \textbf{138}, no.3, 260 (2023)
%doi:10.1140/epjp/s13360-023-03808-6
[arXiv:2208.01179 [hep-th]].

%\cite{Arkhipova:2024iem}
\bibitem{Arkhipova:2024iem}
K.~Arkhipova, L.~Astrakhantsev, N.~S.~Deger, A.~A.~Golubtsova, K.~Gubarev and E.~T.~Musaev,
``Holographic RG flows and boundary conditions in a 3D gauged supergravity,''
Eur. Phys. J. C \textbf{84}, no.6, 560 (2024)
%doi:10.1140/epjc/s10052-024-12932-1
[arXiv:2402.11586 [hep-th]].

\bibitem{Peet:1998wn}
A.~W.~Peet and J.~Polchinski,
``UV / IR relations in AdS dynamics,''
Phys. Rev. D \textbf{59}, 065011 (1999)
%doi:10.1103/PhysRevD.59.065011
[arXiv:hep-th/9809022 [hep-th]].

%\cite{deBoer:1999tgo}
\bibitem{deBoer:1999tgo}
J.~de Boer, E.~P.~Verlinde and H.~L.~Verlinde,
``On the holographic renormalization group,''
JHEP \textbf{08}, 003 (2000)
%doi:10.1088/1126-6708/2000/08/003
[arXiv:hep-th/9912012 [hep-th]].
%\cite{Boonstra:1998mp}

\bibitem{He:2010ye} S.~He, M.~Huang and Q.~S.~Yan,
  ``Logarithmic correction in the deformed $AdS_5$ model to produce the heavy quark potential and QCD beta function'', Phys. Rev. D \textbf{83}, 045034 (2011)
  % doi:10.1103/PhysRevD.83.045034
  [arXiv:1004.1880 [hep-ph]]

\bibitem{Arefeva:2023ter}
  I.~Y.~Aref\textquoteright{}eva, K.~A.~Rannu and P.~S.~Slepov,
  ``Dense QCD in Magnetic Field'',
  Phys. Part. Nucl. Lett. \textbf{20} (2023) no.3, 433-437
  % doi:10.1134/S1547477123030081

\bibitem{Arefeva:2020vae}
  I.~Y.~Aref'eva, K.~Rannu and P.~Slepov,``Holographic model for heavy quarks in anisotropic hot dense QGP with external magnetic field,''JHEP \textbf{07}, 161 (2021)
  %doi:10.1007/JHEP07(2021)161
[arXiv:2011.07023 [hep-th]].

\bibitem{Arefeva:2022avn}
I.~Y.~Aref'eva, A.~Ermakov, K.~Rannu and P.~Slepov,
``Holographic model for light quarks in anisotropic hot dense QGP with external magnetic field,''
Eur. Phys. J. C \textbf{83}, no.1, 79 (2023)
%doi:10.1140/epjc/s10052-022-11166-3
[arXiv:2203.12539 [hep-th]].


\bibitem{Arefeva:2023jjh}
I.~Y.~Aref'eva, A.~Hajilou, K.~Rannu and P.~Slepov,
``Magnetic catalysis in holographic model with two types of anisotropy for heavy quarks,''
Eur. Phys. J. C \textbf{83}, no.12, 1143 (2023)
%doi:10.1140/epjc/s10052-023-12309-w
[arXiv:2305.06345 [hep-th]].
 
\bibitem{Pirner:2009gr}
  H.~J.~Pirner and B.~Galow,
  ``Strong Equivalence of the AdS-Metric and the QCD Running
  Coupling'',
  Phys. Lett. B \textbf{679}, 51-55 (2009)
%doi:10.1016/j.physletb.2009.07.009
  [arXiv:0903.2701 [hep-ph]].

\bibitem{vanRitbergen:1997va}
T.~van Ritbergen, J.~A.~M.~Vermaseren and S.~A.~Larin,
``The Four loop beta function in quantum chromodynamics,''
Phys. Lett. B \textbf{400}, 379-384 (1997)
%doi:10.1016/S0370-2693(97)00370-5
[arXiv:hep-ph/9701390 [hep-ph]].

\bibitem{Arefeva:2018cli}
I.~Aref'eva, K.~Rannu and P.~Slepov,
``Orientation Dependence of Confinement-Deconfinement Phase Transition in Anisotropic Media,''
Phys. Lett. B \textbf{792}, 470-475 (2019)
%doi:10.1016/j.physletb.2019.04.012
[arXiv:1808.05596 [hep-th]].

\bibitem{Arefeva:2020uec}
I.~Y.~Aref'eva, A.~Patrushev and P.~Slepov,
``Holographic entanglement entropy in anisotropic background with confinement-deconfinement phase transition,''
JHEP \textbf{07}, 043 (2020)
%doi:10.1007/JHEP07(2020)043
[arXiv:2003.05847 [hep-th]].

%\cite{Chen:2024jet}
\bibitem{Chen:2024jet}
Y.~Chen, X.~Chen, D.~Li and M.~Huang,
``Deconfinement and chiral restoration phase transition under rotation from holography in an anisotropic gravitational background,''
[arXiv:2405.06386 [hep-ph]].

\bibitem{Gursoy:2017wzz}
  U.~Gursoy, M.~Jarvinen and G.~Nijs,
  ``Holographic QCD in the Veneziano Limit at a Finite Magnetic Field
  and Chemical Potential'',
  Phys. Rev. Lett. \textbf{120}, no.24, 242002 (2018)
  % doi:10.1103/PhysRevLett.120.242002
  [arXiv:1707.00872 [hep-th]]

\bibitem{Bohra:2019ebj}
  H.~Bohra, D.~Dudal, A.~Hajilou and S.~Mahapatra,
  ``Anisotropic string tensions and inversely magnetic catalyzed
  deconfinement from a dynamical AdS/QCD model'',
  Phys. Lett. B \textbf{801}, 135184 (2020)
  % doi:10.1016/j.physletb.2019.135184
  [arXiv:1907.01852 [hep-th]]

\bibitem{Dudal:2021jav}
  D.~Dudal, A.~Hajilou and S.~Mahapatra,
  ``A quenched 2-flavour Einstein\textendash{}Maxwell\textendash{}Dilaton gauge-gravity
  model'',
  Eur. Phys. J. A \textbf{57}, no.4, 142 (2021)
  % doi:10.1140/epja/s10050-021-00461-4
  [arXiv:2103.01185 [hep-th]].

\bibitem{Jain:2022hxl}
P.~Jain, S.~S.~Jena and S.~Mahapatra,
``Holographic confining-deconfining gauge theories and entanglement measures with a magnetic field,''
Phys. Rev. D \textbf{107}, no.8, 086016 (2023)
%doi:10.1103/PhysRevD.107.086016
[arXiv:2209.15355 [hep-th]].


\bibitem{Bohra:2020qom}
H.~Bohra, D.~Dudal, A.~Hajilou and S.~Mahapatra,
``Chiral transition in the probe approximation from an Einstein-Maxwell-dilaton gravity model,''
Phys. Rev. D \textbf{103}, no.8, 086021 (2021)
%doi:10.1103/PhysRevD.103.086021
[arXiv:2010.04578 [hep-th]].

%\cite{Arefeva:2024xmg}
\bibitem{Arefeva:2024xmg}
I.~Y.~Aref'eva, A.~Hajilou, A.~Nikolaev and P.~Slepov,
``Holographic QCD Running Coupling for Light Quarks in Strong Magnetic Field,''
[arXiv:2407.11924 [hep-th]].

\end{thebibliography}
\end{document}